\documentclass[twoside]{article}

\usepackage{qic,epsfig}

\newcommand{\reals}{{\bf R}}

\begin{document}
\setlength{\textheight}{8.0truein}    

\runninghead{A simple proof of the strong subadditivity inequality}
            {Michael A. Nielsen and D\'enes Petz}

\normalsize\textlineskip
\thispagestyle{empty}
\setcounter{page}{1}

\copyrightheading{0}{0}{2005}{000--000}

\vspace*{0.88truein}

\alphfootnote

\fpage{1}

\centerline{\bf A simple proof of the strong subadditivity inequality}

\vspace*{0.37truein}
\centerline{\footnotesize
MICHAEL A. NIELSEN}
\vspace*{0.015truein}
\centerline{\footnotesize\it School of Physical Sciences, The University of
  Queensland}
\baselineskip=10pt
\centerline{\footnotesize\it Brisbane, Queensland 4072, Australia}
\vspace*{10pt}
\centerline{\footnotesize D\'ENES PETZ}
\vspace*{0.015truein}
\centerline{\footnotesize\it 
Department of Mathematical Analysis,
Budapest University of Technology and Economics}
\baselineskip=10pt
\centerline{\footnotesize\it H-1521 Budapest XI, Hungary}
\vspace*{0.225truein}
\publisher{(received date)}{(revised date)}

\vspace*{0.21truein}

\abstracts{
  Arguably the deepest fact known about the von~Neumann entropy, the
  strong subadditivity inequality is a potent hammer in the quantum
  information theorist's toolkit.  This short tutorial describes a
  simple proof of strong subadditivity due to Petz [Rep.  on Math.
  Phys. \textbf{23} (1), 57--65 (1986)].  It assumes only knowledge of
  elementary linear algebra and quantum mechanics.  }{}{}

\vspace*{10pt}

\vspace*{3pt}
\communicate{to be filled by the Editorial}

\vspace*{1pt}\textlineskip    

%
%
\section{Introduction}

The von~Neumann entropy of a density matrix $\rho$ is defined by
$S(\rho) \equiv -\mbox{tr}(\rho \ln \rho)$.  Suppose $\rho_{ABC}$ is a
density matrix for a system with three components, $A, B$ and $C$.
The strong subadditivity inequality states that
\begin{eqnarray}
  \label{eq:ssa}
  S(\rho_{ABC}) + S(\rho_B) \leq S(\rho_{AB})+S(\rho_{BC}),
\end{eqnarray}
where notations like $\rho_B$ denote the appropriate reduced density
matrices.  

The strong subadditivity inequality appears quite mysterious at first
sight.  Some intuition is gained by reexpressing strong subadditivity
in terms of the \emph{conditional entropy} $S(A | B) \equiv
S(\rho_{AB})-S(\rho_B)$.  Classically, when the von Neumann entropy is
replaced by the Shannon entropy function, the conditional entropy has
an intepretation as the average uncertainty about the state of $A$,
given knowledge of the state of $B$~\cite{Cover91a}.  Although this
interpretation is more problematic in the quantum case --- for one
thing, the quantum conditional entropy can be negative!  --- it can
still be useful for developing intuition and suggesting results.  In
particular, we see that strong subadditivity may be recast in the
equivalent form
\begin{eqnarray} \label{eq:ssa-equivalent}
  S(A|BC) \leq S(A|B).
\end{eqnarray}
That is, strong subadditivity expresses the intuition that our
uncertainty about $A$ when $B$ and $C$ are known is not more than when
only $B$ is known.  This intuition is perhaps best viewed as a
mnemonic, due to the problematic interpretation of the conditional
entropy, but may nonetheless be helpful.

Strong subadditivity has many applications in quantum information
theory (see, e.g.,~\cite{Nielsen00a,Ohya04a}).  Our purpose here is
not to discuss these applications, but rather to provide an expository
account of a simple proof of strong subadditivity due to
Petz~\cite{Petz86a} (see also~\cite{Ohya04a}).  In so doing we hope to
help publicise this proof to a wider audience.  The reader looking for
a more comprehensive account in a similar vein to the present paper
should consult~\cite{Petz03a}.

%
%
Our proof strategy is to show that strong subadditivity is implied by
a related result, the \emph{monotonicity of the relative entropy}, and
then to prove this monotonicity result.  The \emph{relative entropy}
between density matrices $\rho$ and $\sigma$ is defined as:
\begin{eqnarray} \label{eq:relative_entropy}
  S(\rho \| \sigma) \equiv \mbox{tr}(\rho \ln \rho - \rho \ln \sigma).
\end{eqnarray}
Roughly speaking, the relative entropy is a measure of the distance
between $\rho$ and $\sigma$.  In particular, it can be shown that
$S(\rho \| \sigma) \geq 0$, with equality if and only if $\rho =
\sigma$.  Be warned, however, that it is not symmetric in $\rho$ and
$\sigma$, and $S(\rho \| \sigma)$ diverges unless the support of
$\rho$ is contained within the support of $\sigma$.  Further
background on the relative entropy may be found
in~\cite{Nielsen00a,Ohya04a}.  The monotonicity of the relative
entropy is the property that discarding part of a composite system
$AB$ can only decrease the relative entropy between two density
matrices $\rho_{AB}$ and $\sigma_{AB}$:
\begin{eqnarray} \label{eq:monotonicity}
  S(\rho_A \| \sigma_A) \leq S(\rho_{AB} \| \sigma_{AB}).
\end{eqnarray}
To see that monotonicity of the relative entropy implies strong
subadditivity, we reexpress strong subadditivity in terms of the
relative entropy, using the identity:
\begin{eqnarray}
  S(B|A) = \ln d_B - 
  S\left( \rho_{AB} \big\| \rho_A \otimes \frac{I_B}{d_b} \right).
\end{eqnarray}
Proving this identity is a straightforward application of the
definitions.  Using this identity we may recast the conditional
entropic form of strong subadditivity, Eq.~(\ref{eq:ssa-equivalent}),
as an equivalent inequality between relative entropies:
\begin{eqnarray} \label{eq:ssa-equivalent-2}
  S\left(\rho_{AB} \big\| \frac{I_A}{d_A} \otimes \rho_{B} \right) \leq
  S\left(\rho_{ABC} \big\| \frac{I_A}{d_A} \otimes \rho_{BC} \right)
\end{eqnarray}
This inequality obviously follows from the monotonicity of the
relative entropy, and thus strong subadditivity also follows from the
monotonicity of the relative entropy.

Strong subadditivity and the monotonicity of the relative entropy have
an interesting and lengthy history, and we will merely note a few
highlights. The reader interested in a more thorough account should
see, e.g., the discussion in~\cite{Ruskai02a,Wehrl78a} and the end
notes to Chapter~11 of~\cite{Nielsen00a}.

The original proof of strong subadditivity was by Lieb and
Ruskai~\cite{Lieb73b}, based on the beautiful concavity results of
Lieb~\cite{Lieb73a}.  Ruskai~\cite{Ruskai04a} has recently given an
elegant exposition along the lines of this original proof.
Monotonicity of the relative entropy was actually proved after strong
subadditivity, by Lindblad~\cite{Lindblad75a}) (see
also~\cite{Uhlmann77a}).  As already noted, our approach to strong
subadditivity and monotonicity is due to Petz~\cite{Petz86a}.
Independently of Petz, Narnhofer and Thirring~\cite{Narnhofer85a}
developed a related approach, based on similar broad ideas, but
differing substantially in the details.  

\section{Background on operator convex functions} 

The only background required for our proofs is a few simple facts from
the theory of operator convex functions.  The reader is referred to
Chapter~5 of~\cite{Bhatia97a} for an introduction to this beautiful
theory.

%
%
Suppose $I \subseteq \reals$ is an interval in the real line, and $f:
I \rightarrow \reals$ is a real-valued function on $I$.  We will
define a corresponding map $f: M_n \rightarrow M_n$, where $M_n$ is
the space of $n \times n$ Hermitian matrices whose spectra lie in $I$.
To define such a map, suppose $D$ is an $n \times n$ diagonal matrix
with real diagonal entries $d_1,\ldots,d_n \in I$.  We define $f(D)$
to be the $n \times n$ diagonal matrix with diagonal entries
$f(d_1),\ldots,f(d_n)$.  Generalizing this definition, if $X$ is any
element of $M_n$ then we can write $X = U D U^\dagger$ for some
unitary $U$ and diagonal matrix $D$.  We define the induced map $f :
M_n \rightarrow M_n$ by $f(U D U^\dagger) \equiv U f(D) U^\dagger$.
More informally, we work in a basis in which $X$ is diagonal, and
apply $f$ to each of the diagonal entries. In cases where $X$ can be
decomposed in many different ways as $X = UD U^\dagger$ it is an easy
exercise to show that $f(X)$ does not depend upon the decomposition
chosen.

%
%
To define operator convexity, we first introduce a partial order on
Hermitian matrices.  Given Hermitian matrices $X, Y \in M_n$ we define
$X \leq Y$ if $Y-X$ is a positive matrix.  We say a function $f : I
\rightarrow \reals$ is \emph{operator convex} if for all $n$, for all
$X, Y \in M_n$, and for all $p \in [0,1]$ we have $f(pX + (1-p)Y) \leq
pf(X)+(1-p)f(Y)$.

%
%
Our later proofs use two simple lemmas about operator convexity, which
we state at the end of this paragraph.  We defer proofs of these
lemmas until after the proof of the monotonicity of relative entropy,
so as to not obscure the simplicity of the ideas used in that proof.

\textbf{Lemma~1:} The function $f(x) = -\ln(x)$ is operator convex.

\textbf{Lemma~2:} If $f$ is operator convex, and $U : V \rightarrow W$
is an isometry (where $\mbox{dim}(V) \leq \mbox{dim}(W)$), then
$f(U^\dagger X U) \leq U^\dagger f(X) U$ for all\footnote{We will
  follow the physicists' convention in often expecting the reader to
  work out from context the domain and range of mappings.  Thus, in
  this example $X$ is a Hermitian matrix on the space $W$, and with a
  spectrum lying within $I$, the domain of $f$.} $X$.

\section{Proof of the monotonicity of the relative entropy} 

To appreciate the ideas used in proving monotonicity, it is helpful to
look at the proof of the analogous classical result.  This states that
for probability distributions $r_{jk}$ and $s_{jk}$ in two variables
we have $\sum_j r_j (\ln r_j-\ln s_j) \leq \sum_{jk} r_{jk} (\ln
r_{jk}-\ln s_{jk})$, where $r_j \equiv \sum_k r_{jk}$ and $s_j \equiv
\sum_k s_{jk}$ are the marginal probability distributions.  This is
easily seen to be equivalent to the inequality $\sum_{jk} r_{jk} \ln
\frac{r_{j}s_{jk}}{r_{jk} s_j} \leq 0$, which may be proved by applying
the calculus result $\ln x \leq x-1$ to the left-hand side, and
showing that the resulting expression vanishes.

The difficulty in the quantum case is that the density matrices
involved may not commute, and this prevents them from being combined
in a single logarithm.  To overcome this difficulty we reexpress the
relative entropy $S(\rho \| \sigma)$ using a linear map on matrices
known as the \emph{relative modular operator}.  In defining this
operator we will assume that $\rho$ and $\sigma$ are invertible; as a
result, our proof of monotonicity of the relative entropy and of
strong subadditivity only applies directly for invertible density
matrices.  The general results follow via a straightforward continuity
argument, which we omit.

To define the relative modular operator, we fix $\rho$ and $\sigma$
and define linear maps on matrices ${\cal L}$ and ${\cal R}$ by ${\cal
  L}(X) \equiv \sigma X$ and ${\cal R}(X) \equiv X \rho^{-1}$, i.e.,
left multiplication by $\sigma$, and right multiplication by
$\rho^{-1}$.  The relative modular operator is defined to be the
product of these linear maps under composition, $\Delta \equiv {\cal
  L} {\cal R}$.  Note that ${\cal L}$ and ${\cal R}$ commute, so we
could equally well have written $\Delta = {\cal R} {\cal L}$.

%
%
Our next step is to define a function $\ln$ on linear maps on
matrices, i.e., to define $\ln({\cal E})$, where ${\cal E}$ is a
linear map on matrices that is strictly positive with respect to the
Hilbert-Schmidt inner product $\langle X, Y\rangle \equiv
\mbox{tr}(X^\dagger Y)$.  To do this we follow the same approach as
described earlier in the section on operator convex functions,
expanding ${\cal E}$ in a diagonal basis as ${\cal E} = \sum_j
\lambda_j {\cal E}_j$, and defining $\ln({\cal E}) \equiv \sum_j
\ln(\lambda_j) {\cal E}_j$.

With this definition, $\ln({\cal L}), \ln({\cal R})$, and
$\ln(\Delta)$ are all defined, i.e., ${\cal L}, {\cal R}$, and
$\Delta$ are all strictly positive with respect to the Hilbert-Schmidt
inner product.  To see that ${\cal L}$ is strictly positive observe
that $\langle X, {\cal L}(X) \rangle = \mbox{tr}(X^\dagger \sigma X) >
0$ for all non-zero $X$.  The proof that ${\cal R}$ is strictly
positive follows similar lines.  Finally, since $\Delta$ is a product
of strictly positive and commuting linear maps on matrices, it follows
that $\Delta$ is strictly positive.

%
%
A little thought shows that $\ln({\cal L})(X) = \ln(\sigma) X$ and
$\ln({\cal R})(X) = -X \ln(\rho)$.  Whatsmore, since ${\cal L}$ and
${\cal R}$ commute, we obtain the beautiful relationship $\ln(\Delta)
= \ln({\cal L})+\ln({\cal R})$.  Some algebra shows that
\begin{eqnarray} \label{eq:critical-rel-ent}
  S(\rho \| \sigma) = \langle \rho^{1/2}, -\ln(\Delta)(\rho^{1/2}) \rangle.
\end{eqnarray}
That is, the relative modular operator has enabled us to combine the
logarithms in the definition of the relative entropy into a single
logarithm, which greatly simplifies analysis.  Using
Eq.~(\ref{eq:critical-rel-ent}) we may rewrite the monotonicity of the
relative entropy in the equivalent form
\begin{eqnarray} \label{eq:critical}
  \langle \rho_A^{1/2}, -\ln(\Delta_A)(\rho_A^{1/2}) \rangle \leq
  \langle \rho_{AB}^{1/2}, -\ln(\Delta_{AB})(\rho_{AB}^{1/2}) \rangle,
\end{eqnarray}
where the first inner product $\langle \cdot, \cdot \rangle$ is on the
space $M(A)$ of matrices acting on $A$, the second inner product is on
the space $M(AB)$ of matrices acting on $AB$, and $\Delta_A(X) \equiv
\sigma_A X \rho_A^{-1}, \Delta_{AB}(X) \equiv \sigma_{AB} X
\rho_{AB}^{-1}$ are the natural relative modular operators on systems
$A$ and $AB$, respectively.

The final step in the proof is to find a linear map on matrices ${\cal
  U}: M(A) \rightarrow M(AB)$ such that: (1) ${\cal U}^\dagger
\Delta_{AB} {\cal U} = \Delta_A$; (2) ${\cal U}(\rho_A^{1/2}) =
\rho_{AB}^{1/2}$; and (3) ${\cal U}$ is an isometry from $M(A)$ to
$M(AB)$.  It is not obvious such a ${\cal U}$ ought to exist.  We
explicitly construct ${\cal U}$ below, but for now we assume ${\cal
  U}$ exists, and investigate the consequences.  Using
Eq.~(\ref{eq:critical}) we rewrite the monotonicity of the relative
entropy as:
\begin{eqnarray}
  & &  \langle \rho_A^{1/2}, -\ln({\cal U}^\dagger \Delta_{AB} 
  {\cal U})(\rho_A^{1/2}) 
  \rangle  \nonumber \\
  &\leq  &
  \langle \rho_{AB}^{1/2}, -\ln(\Delta_{AB})(\rho_{AB}^{1/2}) \rangle. 
\end{eqnarray}
But by Lemma~1 and Lemma~2 on the properties of operator convex
functions we have $-\ln({\cal U}^\dagger \Delta_{AB} {\cal U}) \leq
-{\cal U}^\dagger \ln(\Delta_{AB}) {\cal U}$, and so
\begin{eqnarray}
  & &  \langle \rho_A^{1/2}, -\ln({\cal U}^\dagger \Delta_{AB} 
  {\cal U})(\rho_A^{1/2}) 
  \rangle \nonumber \\
& \leq & \langle \rho_A^{1/2}, -{\cal U}^\dagger \ln(\Delta_{AB})
{\cal U}(\rho_A^{1/2}) \rangle \\
  & = &   \langle {\cal U}(\rho_A^{1/2}), -\ln(\Delta_{AB})
  {\cal U}(\rho_A^{1/2}) \rangle \\
  & = &   \langle \rho_{AB}^{1/2}, -\ln(\Delta_{AB})\rho_{AB}^{1/2} \rangle,
\end{eqnarray}
which completes the proof of monotonicity, provided we can find a
${\cal U}$ satisfying properties (1)-(3).  Based on property (2) a
plausible candidate is ${\cal U}(X) \equiv (X \rho_A^{-1/2} \otimes
I_B) \rho_{AB}^{1/2}$.  With this definition, it is not difficult to
check that ${\cal U}^\dagger(Y) = \mbox{tr}_B( Y \rho_{AB}^{1/2}
(\rho_A^{-1/2} \otimes I_B))$ is the corresponding adjoint operation,
i.e., satisfies $\langle {\cal U}^\dagger(Y), X\rangle = \langle Y,
{\cal U}(X) \rangle$ for all $X \in M(A)$ and $Y \in M(AB)$.  Direct
calculation now shows that ${\cal U}^\dagger \Delta_{AB} {\cal U} =
\Delta_A$ and ${\cal U}^\dagger {\cal U} = {\cal I}_A$, which
completes the list of desired properties, and the proof of
monotonicity.

%
%
This proof of monotonicity highlights the role of the operator
convexity of $f(x) = -\ln(x)$.  If $f$ is any operator convex function
and we define an $f$-relative entropy by $S_f(\rho \| \sigma) \equiv
\langle \rho^{1/2}, f(\Delta)(\rho^{1/2})\rangle$, the same argument
shows that we obtain an analogous monotonicity property.

\section{Proofs of the operator convexity lemmas} 

To prove Lemma~1, we begin with a proof that $f(x) = 1/x$ is operator
convex on $(0,\infty)$.  A key fact used in the proof is that if $X
\leq Y$, then $Z X Z^\dagger \leq Z Y Z^\dagger$ for any choice of
$Z$, i.e., conjugation preserves matrix inequalities.  The proof of
this useful fact is a good exercise in applying the definition of
$\leq$.

To prove the operator convexity of $f(x) = 1/x$, let $X \leq Y$ be
strictly positive Hermitian matrices.  We begin with the special case
$X = I$, where the goal is to prove $(pI + (1-p)Y)^{-1} \leq
pI+(1-p)Y^{-1}$.  Since $I$ and $Y$ commute, this result follows from
the ordinary convexity of the real function $1/x$.

To obtain the general operator convexity from the special case $X =
I$, make the replacement $Y \rightarrow X^{-1/2} Y X^{-1/2}$, which
gives
\begin{eqnarray}
  & & \left( pI+(1-p)X^{-1/2} Y X^{-1/2} \right)^{-1} \nonumber \\
  & \leq & p I+(1-p)(X^{-1/2} Y X^{-1/2})^{-1}.
\end{eqnarray}
Conjugating by $X^{-1/2}$ and doing a little algebra gives the desired
inequality, and concludes the proof that $f(x) = 1/x$ is operator
convex.

The operator convexity of $f(x) = -\ln(x)$ is now established using
the integral representation $-\ln(x) = \int_0^\infty dt \left(
  \frac{1}{x+t} - \frac{1}{1+t} \right)$, from which it follows that
for a strictly positive matrix $X$ we have
\begin{eqnarray} \label{eq:integral-representation}
   -\ln(X) = \int_0^\infty dt ((X+tI)^{-1} - (I+tI)^{-1}).
\end{eqnarray}
Our goal is to show $-\ln(p X + (1-p)Y) \leq -p \ln(X)-(1-p)\ln(Y)$.
From Eq.~(\ref{eq:integral-representation}), this follows if we can
prove $(pX+(1-p)Y + tI)^{-1} \leq p (X+tI)^{-1}+ (1-p) (Y+tI)^{-1}$.
Rewriting the left-hand side as $[p(X+tI)+(1-p)(Y + tI)]^{-1}$ and
applying the operator convexity of $1/x$ gives the desired result,
completing the proof of Lemma~1.

Moving to Lemma~2, note first a simple related result, namely, that
when $U$ maps the space $V$ $\emph{onto}$ $W$, then directly from the
definition of $f(X)$ we obtain $f(U^\dagger X U) = U^\dagger f(X) U$.
This holds true regardless of whether $f$ is operator convex or not.
Lemma~2 requires a stronger hypothesis (the operator convexity of
$f$), and gives rise to an inequality instead of an equality, but has
the advantage that it holds when the range $W'$ of $U$ is a strict
subset of $W$.  Readers familiar with the operator Jensen inequality
(see, e.g.,~\cite{Hansen02a}) may recognize Lemma~2 as a variant of
this result.

To prove Lemma~2, let $P$ be the projector onto $W'$, and $Q \equiv
I-P$ the projector onto the orthocomplement.  As three separate vector
spaces are involved, it is useful to introduce the notations $f_V,
f_W$ and $f_{W'}$ to denote the different ways $f$ can act, e.g.,
$f_V$ takes as input a matrix acting on $V$, and produces as output a
matrix acting on $V$, while $f_W$ takes as input a matrix acting on
$W$, and produces as output a matrix acting on $W$.

Note that $PU = U$, since $P$ projects onto the range of $U$.  As a
result we have $f_V(U^\dagger X U) = f_V(U^\dagger P (PXP) PU)$.  Note
that $PU$ is an isometry from $V$ \emph{onto} $W'$, and since $PXP$
may be regarded as a matrix acting on $W'$, it follows that
$f_V(U^\dagger P (PXP) PU) = U^\dagger P f_{W'}(PXP) PU$.  Summing up,
we have shown that $f_V(U^\dagger XU) = U^\dagger P \, f_{W'}(PXP)
PU$.  A little thought should convince you that to conclude the proof
it will suffice to show that $f_{W'}(PXP) \leq Pf_W(X)P$.  Proving
this inequality now becomes our objective.

We observe that
\begin{eqnarray} \label{eq:inter}
f_{W'}(PXP) & = & Pf_{W}(PXP)P = Pf_W(PXP+QXQ)P,  \nonumber \\
 & & 
\end{eqnarray}
since $f_W(PXP+QXQ) = f_W(PXP)+f_W(QXQ)$ and $Pf_W(QXQ)P = 0$.
Defining a unitary $S \equiv P-Q$ on $W$, and recalling the $P+Q=I$,
we have
\begin{eqnarray}
  \frac{X+SXS^\dagger}{2} = \frac{(P+Q)X(P+Q) + (P-Q)X(P-Q)}{2} = PXP+QXQ,
\end{eqnarray}
for arbitrary $X$.  Applying the operator convexity of $f$ gives
$f_W(PXP+QXQ) \leq (f_W(X)+f_W(SXS^\dagger))/2$, and since
$f_W(SXS^\dagger) = S f_W(X) S^\dagger$ we obtain $f_W(PXP+QXQ) \leq
(f_W(X)+Sf_W(X)S^\dagger)/2 = Pf_W(X)P+Qf_W(X)Q$.  Conjugating by $P$
we obtain $Pf_W(PXP+QXQ)P \leq Pf_W(X)P$.  Combining this inequality
with Eq.~(\ref{eq:inter}) gives $f_{W'}(PXP) \leq Pf_W(X)P$, which, as
noted above, is sufficient to establish Lemma~2.

\nonumsection{Acknowledgements}
\noindent

Thanks to Elliott Lieb and Beth Ruskai for comments on the history of
these inequalities, to Heide Narnhofer for sending us a reprint
of~\cite{Narnhofer85a}, and to Runyao Duan for pointing out some
typographical errors in an early version of this paper.


\end{document}